\documentclass[12pt]{JHEP3}
\usepackage{graphicx}
\usepackage{array}
\usepackage{supertabular}
\usepackage{longtable}

\newcommand{\nn}{\nonumber}
\newcommand{\beqn}{\begin{eqnarray}}
\newcommand{\eeqn}{\end{eqnarray}}
\newcommand{\be}{\begin{equation}}
\newcommand{\ee}{\end{equation}}
\newcommand{\eqn}[1]{(\ref{#1})}

\newcommand{\ba}{\begin{array}{c}}
\newcommand{\bat}{\begin{array}{cc}}
\newcommand{\ea}{\end{array}}

\newcommand{\bi}{\begin{itemize}}
\newcommand{\ei}{\end{itemize}}
\newcommand{\chpt}{$\chi$PT}
\newcommand{\rcht}{R$\chi$T}

\newcommand{\Frac}[2]{\frac{\displaystyle #1}{\displaystyle #2}}

\newcommand{\cO}{{\cal O}}

\newcommand{\gsim}{\stackrel{>}{_\sim}}

\newcommand{\bea}{\begin{eqnarray}}
\newcommand{\eea}{\end{eqnarray}}
\newcommand{\beq}{\begin{equation}}
\newcommand{\eeq}{\end{equation}}

\newcommand{\bear}{\begin{eqnarray}}
\newcommand{\eear}{\end{eqnarray}}
\newcommand{\mF}{\mathcal{F}}

\newcommand{\ket}{\,\rangle}
\newcommand{\bra}{\langle \,}

\title{The vector form factor at the next-to-leading order in $1/N_C$: chiral couplings
$\mathbf{L_{9}(\mu)}$ and $\mathbf{C_{88}(\mu)}-\mathbf{C_{90}(\mu)}$}

\author{Antonio Pich,$^a$ Ignasi Rosell$^{ab}$ and Juan Jos\'e Sanz-Cillero$^c$ \\
$^a$Departament de F\'\i sica Te\`orica, IFIC, Universitat de Val\`encia - CSIC \\
Apt. Correus 22085, E-46071 Val\`encia, Spain \\
$^b$Departamento de Ciencias F\'\i sicas, Matem\'aticas y de la Computaci\'on, \\
Universidad CEU Cardenal Herrera, c/ Sant Bartomeu 55, \\
E-46115 Alfara del Patriarca, Val\`encia, Spain \\
$^c$ Istituto Nazionale di Fisica Nucleare INFN, Sezione di Bari,\\
Via Orabona 4, I-70126 Bary, Italy \\
E-mail: \email{Antonio.Pich@uv.es,rosell@uch.ceu.es,cillero@ifae.es}}

\abstract{
Using the Resonance Chiral Theory Lagrangian, we perform a calculation
of the vector form factor of the pion at the next-to-leading order (NLO)
in the $1/N_C$ expansion. Imposing the correct QCD short-distance
constraints,  one fixes the amplitude
in terms of the pion decay constant $F$ and resonance masses.
Its low momentum expansion determines then the corresponding
$\cO(p^4)$ and $\cO(p^6)$  low-energy chiral couplings at NLO, keeping control of their
renormalization scale dependence. At $\mu_0=0.77$~GeV, we obtain
$L_{9}(\mu_0) = (7.9 \pm 0.4)\cdot 10^{-3}$ and
$C_{88}(\mu_0)-C_{90}(\mu_0)=(-4.6 \pm 0.4)\cdot 10^{-5}$.}


\preprint{BARI$-$TH/635$-$10 \\ FTUV/10$-$1126 \\ IFIC/10$-$49\\}

\begin{document}

\section{Introduction}

Effective field theories (EFT) are
nowadays the standard tool to investigate
the low-energy dynamics of Quantum Chromodynamics
(QCD). In particular, the chiral symmetry is a crucial ingredient for
the understanding of the light quark interactions.
The dynamics of the pseudo-Goldstone bosons from the spontaneous symmetry breaking
is provided by the corresponding EFT, Chiral Perturbation
Theory~(\chpt), with a perturbative expansion
in powers of light quark masses and external
momenta~\cite{Weinberg,ChPTp4}.
This allows a systematic description of the long-distance
regime of QCD, at energies below the lightest resonance mass.
The precision required in present
phenomenological applications makes necessary to include
corrections of $\cO(p^6)$. While many two-loop \chpt\ calculations
have been already carried out~\cite{ChPTp6},
the large number of unknown low-energy constants (LECs) appearing
at this order puts a clear limit to the achievable accuracy.
The determination of these \chpt\ couplings is compulsory
to achieve further progress in our understanding of strong interactions
at low energies.

In the resonance region, $E\sim M_R$, the chiral counting breaks down
and the new heavier degrees of freedom --the resonances--
have to be explicitly incorporated into the theory.
A suitable alternative is then provided by the $1/N_C$ expansion
in the limit of a large number of colours,
$N_C\to\infty$~\cite{NC,MHA,polychromatic}.
Assuming confinement, the strong
dynamics is given at large $N_C$ by tree-level diagrams with an
infinite number  of possible hadronic exchanges.
This corresponds to the tree approximation
of some local  Lagrangian, being meson loops suppressed
by higher powers of $1/N_C$~\cite{NC}.
Resonance Chiral Theory (\rcht) provides an appropriate
framework to incorporate these massive
mesonic states within a chiral invariant phenomenological
Lagrangian~\cite{RChTa,RChTb,RChTc}.
The operators of the \rcht\ action are constructed such  that they
remain unchanged under flavour  transformations $U(3)_L\otimes U(3)_R$.
After integrating out the heavy
fields,   the \chpt\ Lagrangian is recovered at low energies with
explicit values of the chiral LECs in terms of resonance parameters.
The short-distance properties of QCD impose stringent constraints
on the \rcht\ couplings  and  provide important information
for the extraction of the low-energy \chpt\ parameters.
The amplitudes are thus enforced to follow the known high-energy QCD
behaviour, introducing in the long-distance description
important information from the underlying theory~\cite{MHA,polychromatic}.

Clearly, we cannot determine at present the infinite number of meson
couplings which characterize the large--$N_C$ Lagrangian.
However, one can perform  useful
approximations in terms of a finite number of meson fields. Truncating
the infinite tower of mesons to the lowest resonances with
$0^{-+}$, $0^{++}$, $1^{--}$ and $1^{++}$ quantum numbers,
one gets a very successful
prediction for the $\cO(p^4)$ \chpt\ couplings
at large $N_C$~\cite{polychromatic}.
Already at this level the comparison with experimental determinations
of the $\cO(p^4)$ chiral couplings shows a remarkable agreement.
Some $\cO(p^6)$ LECs have been also
estimated in this way, by studying   appropriate sets of
Green functions (see ref.~\cite{RChTc} and references therein).
All the required terms in the \rcht\ Lagrangian that may contribute
to the $\cO(p^6)$ LECs at LO in $1/N_C$ were classified in
ref.~\cite{RChTc}.

Since chiral loop corrections are of next-to-leading order (NLO)
in the $1/N_C$ expansion, the large--$N_C$ determination of the
LECs is unable to control their renormalization-scale dependence.
%
%
First analyses of resonance loop contributions to
the running of $L_{10}(\mu)$ and $L_9(\mu)$ were attempted
in refs.~\cite{CP:02} and~\cite{RSP:05}, respectively.
In spite of all the complexity associated with the still not
so well understood renormalization
of \rcht\ \cite{RSP:05,RPP:05,RChT-EoM,saturation,natxo-tesis,vectors-Kampf},
these pioneering calculations showed the potential
predictability at the NLO in $1/N_C$.

Using dispersion relations we can avoid the technicalities associated with
the renormalization procedure~\cite{natxo-tesis,L8-nlo,L10-nlo}.
This allows one to understand the underlying
physics in a much more transparent way.
Still, a fully equivalent diagrammatic calculation is possible, although the
derivation and presentation is slightly more
cumbersome~\cite{CP:02,RSP:05,L8-Trnka}.
In particular, the subtle cancellations among many unknown renormalized
couplings found in ref.~\cite{RSP:05} and the relative simplicity of
the final result can be better understood in terms of the imposed
short-distance constraints within the dispersive approach.
Following these ideas we determined, up to NLO in $1/N_C$, the couplings
$L_8(\mu)$ and $C_{38}(\mu)$ in ref.~\cite{L8-nlo}
and $L_{10}(\mu)$ and $C_{87}(\mu)$ in  ref.~\cite{L10-nlo}. In this article we present  the study of
the vector form factor (VFF) of the pion, which allows us to
estimate the \chpt\ coupling $L_{9}(\mu)$ and the $\cO(p^6)$ combination
$ C_{88}(\mu)-C_{90}(\mu) $ up to NLO in $1/N_C$.

In order to establish the notation, the \rcht\ Lagrangian is introduced
in the next section. The analysis of the VFF in the resonance region
is performed in section~\ref{sec:VFF},
while section~\ref{sec:L9} contains the determination of $L_{9}(\mu)$
and $C_{88}(\mu)-C_{90}(\mu)$.
A summary of our results is finally given in section~\ref{sec:conclusions}.
In order to ease the reading of the text,
we have shifted the technical details on the calculation of the spectral
function,  the full VFF and the chiral coupling expressions to
the Appendices.

\section{The Lagrangian}

We will  adopt the Single Resonance Approximation (SRA),
where just the lightest resonances with non-exotic quantum numbers
are considered.\footnote{
In ref.~\cite{Pade}, it has been argued that large discrepancies
may occur between the values of the masses and couplings
of the full large--$N_C$ theory and those from descriptions with a finite number
of resonances. Even in this case, it is found that one can obtain safe
determinations of the LECs as far as one is able to construct a good
interpolator that reproduces the right asymptotic behaviour at  low and high energies.
Further issues related to the truncation of the spectrum to a finite number of resonances are
discussed in ref.~\cite{inf-res}.}
On account of the large-$N_C$ limit, the mesons are put together
into $U(3)$ multiplets. Hence, our degrees of freedom are the
pseudo-Goldstone bosons (the lightest pseudoscalar mesons)
along with massive multiplets of the type $V(1^{--})$, $A(1^{++})$,
$S(0^{++})$ and $P(0^{-+})$. With them, we construct the most general
action that preserves chiral symmetry. Since we are interested in
determining the \chpt\ low-energy constants and the study of the
short-distance behaviour,  the chiral limit will be taken all along the
paper.  No information is lost as the chiral LECs  are independent
of the light quark masses.

Resonance Chiral Theory must satisfy the high-energy behaviour dictated
by QCD. To comply with this requirement we will only consider operators
constructed with chiral tensors of $\cO(p^2)$; interactions with
higher-order chiral tensors tend to violate the  asymptotic
short-distance behaviour prescribed by
QCD~\cite{polychromatic,saturation}.
Likewise, it has been shown in some cases that
resonance operators  with higher number
of derivatives can be simplified into
terms with less derivatives, terms without resonances
and operators that contribute to other
hadronic amplitudes,   by means of the equations of motion and
convenient meson field
redefinitions~\cite{RChTa,RChTc,RSP:05,RPP:05,RChT-EoM,L8-Trnka}.

The different terms in the Lagrangian can be classified by their number of resonance fields:
\begin{eqnarray} \label{lagrangian}
\mathcal{L}_{R\chi T}&=&\mathcal{L}_G \,+\,\sum_{R_1}\mathcal{L}_{R_1}
\,+\,\sum_{R_1,R_2}\mathcal{L}_{R_1R_2}
\, + \, ... \,\,\,  ,
\end{eqnarray}
where the dots denote operators with three
or more resonance fields, and the indices $R_i$ run over
all different resonance multiplets, $V$, $A$, $S$ and $P$.
The term with only pseudo-Goldstone bosons is given by~\cite{ChPTp4}
\begin{eqnarray}
\mathcal{L}_{G} &=& \frac{F^2}{4} \bra u_\mu u^\mu + \chi_+ \ket \, .
\end{eqnarray}
The second term in eq.~\eqn{lagrangian} corresponds
to the operators with one massive resonance~\cite{RChTa},
\begin{eqnarray}
\mathcal{L}_V &=&\, \frac{F_V}{2\sqrt{2}} \bra V_{\mu\nu} f^{\mu\nu}_+ \ket \,+\, \frac{i\, G_V}{2\sqrt{2}} \bra V_{\mu\nu} [u^\mu, u^\nu] \ket \, , \nonumber \\
\mathcal{L}_A &=&\, \frac{F_A}{2\sqrt{2}} \bra A_{\mu\nu} f^{\mu\nu}_- \ket\, , \nonumber\\
%
\mathcal{L}_S &=&\, c_d \bra S u_\mu u^\mu\ket\,+\,c_m\bra S\chi_+\ket\,  \, , \phantom{\frac{1}{2}}  \nonumber \\
\mathcal{L}_P &=&\, i\,d_m \bra P \chi_- \ket\, .\phantom{\frac{1}{2}} \label{P}
\end{eqnarray}
The Lagrangian $\mathcal{L}_{R_1R_2}$ contains the kinetic resonance
terms and the remaining operators with
two resonance fields~\cite{RChTa,RChTc,RSP:05}. We show only the
terms that contribute to the vector form factor of the pion,
taking into account
that here we just consider the lowest-mass two-particle absorptive channels,
with two pseudo-Goldstone bosons or one pseudo-Goldstone and one resonance.
In the energy range we are interested in, exchanges of two heavy
resonances are kinematically suppressed. Hence, the
relevant operators are
\begin{eqnarray}
%
%
%
%
%
\Delta \mathcal{L}_{SA}&=&\, \lambda^{SA}_1 \bra \{\nabla_\mu S, A^{\mu\nu} \} u_\nu \ket  
 \, ,\phantom{\frac{1}{2}}  \nonumber\\
 \Delta \mathcal{L}_{SP}&=&\,\lambda^{SP}_1 \bra u_\alpha \{\nabla^\alpha S, P \} \ket \,,\phantom{\frac{1}{2}} \nonumber \\
%
%
\Delta \mathcal{L}_{PV}&=&\, i \lambda^{PV}_1\bra [\nabla^\mu P,V_{\mu\nu} ] u^\nu \ket  
\, , \phantom{\frac{1}{2}} \nonumber \\
%
%
\Delta \mathcal{L}_{VA}&=&\,  
i \lambda^{VA}_2 \bra  [ V^{\mu\nu}, A_{\nu\alpha} ] h^\alpha_\mu \ket \,+\, i \lambda^{VA}_3 \bra  [ \nabla^\mu V_{\mu\nu}, A^{\nu\alpha} ] u_\alpha \ket  \phantom{\frac{1}{2}} \nonumber \\
&&\,+\, i \lambda^{VA}_4 \bra  [ \nabla_\alpha V_{\mu\nu}, A^{\alpha\nu} ] u^\mu \ket \,+\, i \lambda^{VA}_5 \bra  [ \nabla_\alpha V_{\mu\nu}, A^{\mu\nu} ] u^\alpha \ket  \phantom{\frac{1}{2}} \,.
%
\end{eqnarray}
All coupling constants are real,
the brackets $\langle ... \rangle$ denote a trace of the corresponding
flavour matrices, and the standard  definitions for the
$u^\mu$, $\chi_\pm$, $f_\pm^{\mu\nu}$ and $h^{\mu\nu}$ chiral tensors
of pseudo-Goldstones  are provided in refs.~\cite{RChTa,RChTc}.

Our Lagrangian $\mathcal{L}_{R\chi T}$ satisfies the $N_C$ counting rules for
a theory with $U(3)$ multiplets. Therefore, only operators that have one trace
in the flavour space are considered. Note that  local terms with two traces
in flavour space, which are of NLO in $1/N_C$, cannot contribute at
tree-level to the VFF because the final two-pion state has isospin $I=1$.
The different fields,
masses and momenta are of $\cO(N_C^0)$ in the $1/N_C$ expansion.
Taking into account the interaction terms, one can check that
$F,\,F_V,\,G_V,\,F_A,\,c_d,\,c_m $ and $d_m$ are   $\cO(\sqrt{N_C})$ and
the $\lambda_i^{R_1R_2}$  are $\cO(N_C^0)$
. The mass dimension of these parameters is
$[F]=[F_V]=[G_V]=[F_A]=[c_d]=[c_m]=[d_m]=
E$ and $[\lambda_i^{R_1R_2}]=E^0$.
%

Note that the $U(3)$ equations of motion have been used in order to reduce the number
of operators. For instance, terms like $\langle P\,\nabla_\mu u^\mu \rangle$ are
not present in eq.~(\ref{P}), since they can be transformed
into operators that, either have been already considered,
or contain a higher number of mesons by means of the equations of motion
and convenient meson field redefinitions~\cite{RChTa}.

The \rcht\ Lagrangian~\eqn{lagrangian} contains a large number
of unknown coupling constants. However, as we will see in the next
section, the short-distance QCD constraints allow
us to determine many of them. In the observable at hand and with our
assumptions, we initially have ten couplings or combinations of them
($F$, $F_V$, $G_V$, $F_A$, $c_d$, $\lambda^{SA}_1$, $\lambda^{SP}_1$,
$\lambda^{PV}_1$, $-2 \lambda_2^{\mathrm{VA}} + \lambda_3^{\mathrm{VA}}$ and
 $2 \lambda_2^{\mathrm{VA}} -2 \lambda_3^{\mathrm{VA}}
+ \lambda_4^{\mathrm{VA}} + 2 \lambda_5^{\mathrm{VA}}$)
and four resonance masses ($M_V$, $M_A$, $M_S$ and $M_P$).
As we will see in section~\ref{sec:VFF}, after imposing a good
short-distance behaviour of this observable, the number of parameters reduces
to three couplings  ($F$, $G_V$ and $F_A$) and three masses
($M_V$, $M_A$ and $M_S$).
The Weinberg sum-rules associated with the left--right correlator \cite{SR}
allow us to further reduce the number of inputs;
the amplitude is finally determined in terms of just $F$ and the three
masses $M_V$, $M_A$ and $M_S$.
The role of the information coming from the
underlying theory is thus fundamental.

\section{The vector form factor of the pion} \label{sec:VFF}

Our observable is defined through the two pseudo-Goldstone matrix element of the vector current:
\begin{eqnarray}
\bra \pi^+(p_1)\, \pi^-(p_2)\,| \,\frac{1}{2}\left( \bar{u}\gamma^\mu u - \bar{d}\gamma^\mu d \right)|0 \ket & = & \mathcal{F}(s) \, (p_1-p_2)^\mu\, ,
\end{eqnarray}
where $s\equiv (p_1+p_2)^2$. At very low energies,
$\mathcal{F}(s)$ has been studied within the $\chi$PT framework
up to $\cO(p^6)$~\cite{ChPTp4,VFF_ChPT}. R$\chi$T and the $1/N_C$
expansion have also been used to determine $\mathcal{F}(s)$ at the $\rho$
meson peak, including appropriate resummations of subleading
logarithms from two pseudo-Goldstone channels~\cite{preVFF1,preVFF2}.
A first systematic study of the VFF
at NLO in $1/N_C$ was performed in ref.~\cite{RSP:05}.
Although the general structure  was well established there,
the present article answers and solves three important questions
raised  in that previous paper:

\begin{itemize}

\item{}
 In ref.~\cite{RSP:05}  only operators with at most one resonance
field were included (except for the kinetic resonance terms)~\cite{RChTa}.
However,  as suggested in the appendix~C of
that article, this assumption is not really
justified and leads to problems with the asymptotic
short distance behaviour.
In the present paper, we have considered  all the operators needed
to describe the   absorptive cuts with
two chiral pseudo-Goldstones and those with one pseudo-Goldstone
and one resonance,
being higher thresholds with two resonances highly
suppressed in the energy region that we consider~\cite{L10-nlo}.

\item{}
 Due to this first issue, in ref.~\cite{RSP:05} the logarithmic part of
$\mathcal{F}(s)$ was badly behaved
at  high energies. It was not possible to enforce a vanishing  form factor
at $s\to \infty$ without the inclusion
of new hadronic operators in the leading Lagrangian.
The inclusion of those terms  in the present article will
allow us to recover the expected high-energy dependence for the VFF
in QCD~\cite{brodsky-lepage}.

\item{}
 The final result of ref.~\cite{RSP:05}
 contained the unknown \rcht\ couplings $\widetilde{L}_9$ and
$\widetilde{C}_{88}-\widetilde{C}_{90}$,  which are
the analogous  ones to the \chpt\ LECs $L_9$ and $C_{88}-C_{90}$.
In the present work, they are fully determined by means of the
high-energy matching with QCD~\cite{saturation}.

\end{itemize}

\begin{figure}
\begin{center}
\includegraphics[scale=0.8]{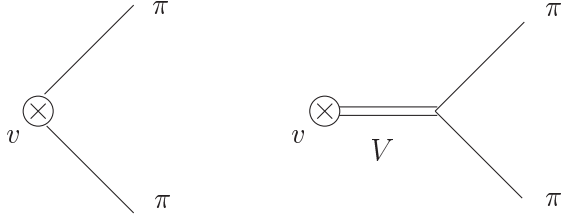}
\caption{\label{vgg}
Tree-level contributions to the vector form factor of the pion.
A single line stands for a pseudo-Goldstone boson while
a double line indicates a resonance.}
\end{center}
\end{figure}

Within Resonance Chiral Theory the diagrams contributing to the VFF at leading order in $1/N_C$ are shown in figure~\ref{vgg}. They generate the result
\begin{eqnarray}
\mathcal{F}_{R \chi T} \left(s\right) &=&  1 + \frac{F_VG_V}{F^2} \frac{s}{M_V^2 - s} \,.
\end{eqnarray}
Considering that the form factor is constrained to be zero at infinite momentum transfer~\cite{brodsky-lepage}, the vector couplings should satisfy
\begin{eqnarray}
F_V G_V &=& F^2 \,, \label{constraintLO}
\end{eqnarray}
which implies
\begin{eqnarray}
\mathcal{F}_{R \chi T} \left(s\right) &=&  \frac{M_V^2}{M_V^2 - s} \,. \label{VFFLO}
\end{eqnarray}

\begin{figure}
\begin{center}
\includegraphics[scale=0.19]{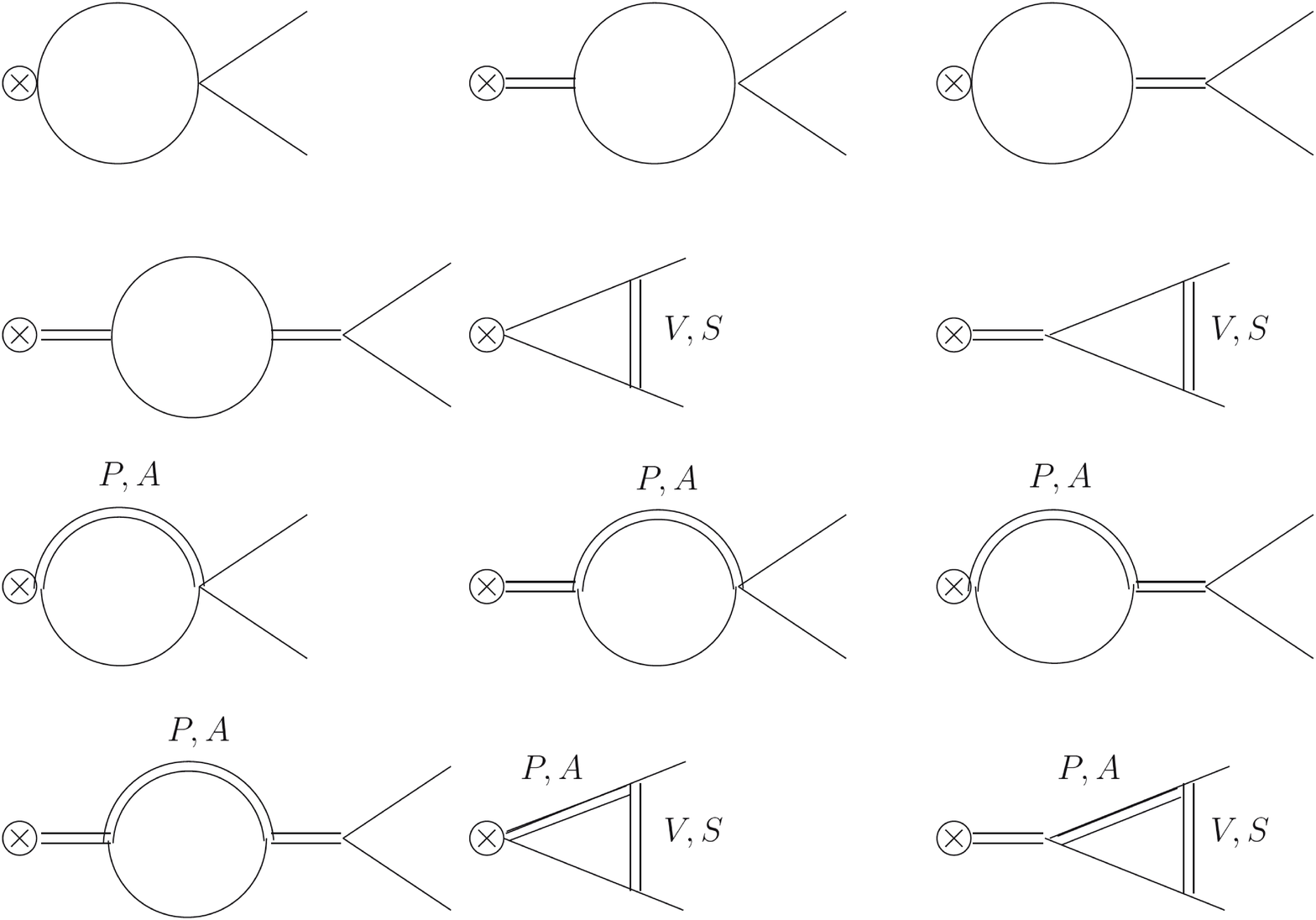}
\caption{\label{feynmanNLO}
One-loop contributions to the vector form factor of the pion with absorptive cut.
A single line stands for a pseudo-Goldstone boson while
a double line indicates a resonance.}
\end{center}
\end{figure}

The subleading corrections can be calculated by means of dispersive
relations. Once the one-loop absorptive parts of
$\mathcal{F}_{R \chi T} \left(s\right)$ are known,
one can reconstruct  the full form factor up to appropriate
subtraction terms.
We can separate then the leading and subleading parts
of the amplitude in the form
\begin{eqnarray}
\mathcal{F}_{R \chi T} \left(s\right) &=&
\, \Frac{M_V^2}{M_V^2-s} \,\,+\, \,
\mF(s)_{_{\rm NLO}}\, ,
\label{eq.VFFNLO}
\end{eqnarray}
with $\mF(s)^{^{\rm NLO}}$ containing the one-loop contribution and the
subleading part $\delta_{_{\rm NLO}}$
of the resonance coupling combination
$F_VG_V/F^2=1+\delta_{_{\rm NLO}}$ (for details see appendix \ref{app.diagrams}):
\begin{eqnarray}
\mF(s)_{_{\rm NLO}}\,\,=\,\, \delta_{_{\rm NLO}} \,\Frac{s}{M_V^2-s}
\,\,+\,\, \mF^{1\ell}(s)\, . \label{deltaNLO}
\end{eqnarray}
The explicit form for the subtracted one-loop
amplitude $\mF^{1\ell}(s)$
can be found in Appendices~\ref{app.disp-rel} and \ref{app.F1loop},
being fully determined by the spectral function Im$\mF(s)$ through a
once-subtracted dispersion relation.
It vanishes at $s=0$ and has no contribution to the real part of the
pole at $s=M_V^2$.
The subleading correction to the couplings, $\delta_{_{\rm NLO}}$, is
fixed by means of the high-energy matching after demanding that it
cancels the bad behaviour of
$\mF^{1\ell}(s)=\delta_{_{\rm NLO}}+\cO(s^{-1})$
when $s\to\infty$.
Furthermore, the NLO term $\mF(s)_{_{\rm NLO}}$ can be neatly
separated into its different contributions from the
various two-meson absorptive channels $\mF(s)_{_{\rm NLO}}|_{m_1,m_2}$,
given by the corresponding $\mF^{1\ell}(s)|_{m_1,m_2}$
and the consequent $\delta_{_{\rm NLO}}|_{m_1,m_2}$. These details are relegated to
Appendices~\ref{app.ImF} and \ref{app.F1loop}.

Although in this article we follow the procedure of
refs.~\cite{L8-nlo,L10-nlo}, our results  can be also derived in an utterly
equivalent way through a  Feynman diagram computation  and
the standard renormalization procedure. 
This derivation is slightly more complex and its detailed  explaination
is relegated to appendix~\ref{app.diagrams}.

We will consider only the effects of absorptive loops with
two pseudo-Goldstones
($\pi\pi$) or with one pseudo-Goldstone and a resonance ($R\pi$).
Two-resonance channels $RR'$ have their thresholds at
$(M_R+M_R')^2\gsim 2$~GeV$^2$ and their impact on the LEC determination
is expected to be negligible~\cite{L10-nlo}.
Taking this into account, we extract
our \rcht\ form factor through the following
short-distance matching procedure:

\begin{enumerate}

\item Determine the spectral function of the considered
 absorptive cuts ($\pi\pi$ and $R\pi$).
The full expressions are shown in eqs.~(\ref{Impipi1}), (\ref{ImPpi1})
and (\ref{ImApi1}) of appendix~\ref{app.ImF}.

\item
We demand Im$\mF(s)$ to be well-behaved
at high energies, {\it i.e.}, it must vanish when $s\to\infty$.
In the present work,  we will actually
impose this constraint channel by channel, {\it i.e.}, we will demand
that each separate two-meson cut    Im$\mF(s)|_{m_1,m_2}$   vanishes
at $s\to\infty$.  For spin--0 mesons this must be so as
its one-loop contribution to the spectral function is essentially
its VFF at LO (which vanishes at infinite momentum)
times the partial-wave scattering amplitude at LO (which
is upper bounded).  For    higher spin resonances
the derivation is more cumbersome as the Lorentz structure
allows for the proliferation of form factors and
the unitarity relations are not that simple. Still, in many situations
it has been already found that  amplitudes with
massive spin--1 mesons as final states must  go to zero
at high energies even faster, due to the presence of extra powers
of momenta in the unitarity relations coming from intermediate
longitudinal polarizations~\cite{L10-nlo}.
In summary, we will assume Im$\mF(s)|_{m_1,m_2}\to 0$ when
$s\to\infty$ for every absorptive two-meson cut under consideration,
regardless of the spin of the intermediate mesons.

In the case of the $\pi\pi$ cut we have found two constraints, which are consistent with the literature,
\begin{equation}
F_V G_V\,=\, F^2 \,, \qquad \qquad    3\,G_V^2+2\,c_d^2 \,=\, F^2 \,, \label{constraintpipi}
\end{equation}
where the first one coincides with eq.~(\ref{constraintLO}), that is, with the constraint obtained with the vector form factor at leading-order~\cite{RChTb}.
The second one was derived in ref.~\cite{JJGuo}  from the LO
$\pi\pi$ scattering amplitude.
It is interesting to remark that the $c_d=0$ limit of this second relation,
$G_V=F/\sqrt{3}$, has been obtained recently from a study of
$\tau^- \to P^- \, \gamma\, \nu_\tau$ decays ($P=\pi,K$)~\cite{guo-roig}.
We have used these constraints to fix $F_V$ and $c_d^2$.

For the $P\pi$ cut, the only possible solution is to kill
the whole  contribution by means of
\begin{eqnarray}
\lambda_1^{\mathrm{PV}}\,=\,0\,, \label{constraintPpi}
\end{eqnarray}
which is consistent with the large-$N_C$ constraint from
the vector form factor into $P\pi$, studied in ref.~\cite{L10-nlo}.

The analysis of the $A\pi$ cut leads to more than
one real solution. We have chosen  the solutions consistent with
previous works~\cite{natxo-tesis,L10-nlo},
where the NLO contributions in $1/N_C$ to the
$\Pi_{VV}(s)$ correlator coming from
tree-level form factors to resonance fields were studied:
\begin{eqnarray}
&&
  -2 \lambda_2^{\mathrm{VA}} + \lambda_3^{\mathrm{VA}}=
0 \,, \phantom{\frac{1}{2}}
\qquad
- \lambda_3^{\mathrm{VA}} + \lambda_4^{\mathrm{VA}}
+ 2 \lambda_5^{\mathrm{VA}} = \frac{F_A}{F_V} \,,\nonumber
\\
&&
\lambda_1^{\mathrm{SA}}=
-\frac{F_A \,G_V\left(M_A^2 - 4\,M_V^2\right)}{3\sqrt{2} M_A^2 c_d\,F_V}
\,.
\label{constraintApi}
\end{eqnarray}
The first two constraints, in the first line, come from the
analysis of the $A\pi$ vector form-factor. The last relation
with  $\lambda_1^{SA}$  is then
needed to make Im$\mF(s)|_{A\pi}\to 0$ for $s\to\infty$.

After imposing  the relations
(\ref{constraintpipi}), (\ref{constraintPpi}) and (\ref{constraintApi})
the spectral functions can be expressed in terms of $G_V$, $F_A$, $F$
and masses, as shown in eqs.~(\ref{Impipi}), (\ref{ImPpi})
and (\ref{ImApi}).    

\item The spectral function is now ready  for the once-subtracted
dispersion relation provided in the appendix \ref{app.disp-rel} in eq.~(\ref{eq.F1loop}),
which allows to reconstruct the
full form factor up to the pole position at $s=M_V^2$ and
the real part of its residue.

\item
Finally, we impose that the whole  $\mathcal{F}_{R \chi T} (s)$
vanishes at short distances --not only its imaginary part--.
This fixes the real part of the residue at $s=M_V^2$ and,
consequently, the NLO correction
$\delta_{\mathrm{NLO}}$ in eq.~(\ref{deltaNLO}).
%
%
In order to ease the reading of the manuscript, the complicated
expressions for the well-behaved contributions to
the different channels are provided in appendix~\ref{app.F1loop},
in eqs.~(\ref{VFFNLOpipi}), (\ref{VFFNLOPpi}) and (\ref{VFFNLOApi}).

\end{enumerate}

\section{The chiral couplings $L_{9}(\mu)$ and $C_{88}(\mu)-C_{90}(\mu)$} \label{sec:L9}

The low-momentum expansion of $\mathcal{F}(s)$ is determined by $\chi$PT~\cite{ChPTp4,VFF_ChPT}. The corresponding expression in the chiral limit reads
\begin{equation}
\begin{array}{rl}
&\mathcal{F}_{\chi PT} \left(s\right) =  1 + \displaystyle\frac{2\,s}{F^2} \left\{L_9(\mu) +\frac{\Gamma_9}{32\pi^2}  \left( \frac{5}{3}-\log \frac{-s}{\mu^2} \right) \right\} \nonumber \\& \qquad
 -\, \displaystyle\frac{4\,s^2}{F^4} \left\{ C_{88}(\mu) - C_{90}(\mu) -\frac{\Gamma_{88}^{(L)}-\Gamma_{90}^{(L)} }{32\pi^2}
\left( \frac{5}{3}-\log \frac{-s}{\mu^2} \right) +\mathcal{O}\!\left(N_C^{0}\right) \right\} +
 \mathcal{O}\! \left(s^3\right)\, ,
 \end{array}\label{VFFChPT}
\end{equation}
with~\cite{ChPTp4,ChPTp6}
\begin{equation}
\Gamma_{9}\, =\, \frac{1}{4}, \qquad \qquad
\Gamma_{88}^{(L)}-\Gamma_{90}^{(L)} \,=\,
-\frac{2L_1}{3}+\frac{L_2}{3}-\frac{L_3}{2}+\frac{L_9}{4}\,. \label{matching}
\end{equation}
The couplings $F^2$, $L_{9},\,C_{88}/F^2$
and $C_{90}/F^2$ are of $\cO(N_C)$, while
$\Gamma_{9}$, $\Gamma_{88}^{(L)}/F^2$
and $\Gamma_{90}^{(L)}/F^2$ are of $\cO(N_C^0)$ and represent a NLO effect.

The low-energy expansion of eqs.~(\ref{VFFLO}) and (\ref{eq.VFFNLO}),
obtained, respectively, within Resonance Chiral Theory at leading-order and at
next-to-leading order in the $1/N_C$ expansion, allows to determine the chiral couplings $L_9$ and $C_{88}-C_{90}$ at LO and at NLO.

\subsection{The large-$N_C$ limit}

At leading-order in $1/N_C$, eq.~(\ref{VFFChPT}) becomes
\begin{equation}
\mathcal{F}_{\chi PT} \left(s\right) \,=\,  1 + \frac{2\,s}{F^2} \left\{ L_9 + \mathcal{O}\!\left(N_C^{0}\right) \right\}
 - \frac{4\,s^2}{F^4} \left\{ C_{88} - C_{90} + \mathcal{O}\!\left(N_C\right)\right\} + \mathcal{O} \!\left(s^3\right) \,. \label{ChPTLO}
\end{equation}
Within R$\chi$T in the large-$N_C$ limit, eq.~(\ref{VFFLO}) can be now expanded at low energies:
\begin{equation}
\mathcal{F}_{R \chi T} \left(s\right) \,=\,  \frac{M_V^2}{M_V^2 - s}  \,=\, 1 + \frac{s}{M_V^2} + \frac{s^2}{M_V^4} + \mathcal{O}\! \left(s^3\right)   \,. \label{expansionLO}
\end{equation}
The matching between (\ref{ChPTLO}) and (\ref{expansionLO}) fixes $L_9$ and $C_{88}-C_{90}$ at LO~\cite{RChTb,RChTc},
\begin{equation}
L_9\,=\,\frac{F^2}{2M_V^2} \,, \qquad \qquad C_{88}-C_{90}\,=\, -\frac{F^4}{4M_V^4} \,. \label{L10LO}
\end{equation}

\subsection{$L_{9}(\mu)$ and $C_{88}(\mu)-C_{90}(\mu)$ at NLO}

Following the same steps as before, let us determine the related $\mathcal{O}(p^4)$ and $\mathcal{O}(p^6)$ low-energy constants by matching eq.~(\ref{VFFChPT}) and the low-energy expansion of eq.~(\ref{eq.VFFNLO}),
\begin{eqnarray}
\mathcal{F}_{R \chi T} \left(s\right) &=& 1
+ \displaystyle\frac{2s}{F^2} \left\{ \frac{F^2}{2M_V^2}
+  \bar\xi^{(2)}
+\frac{\Gamma_9}{32\pi^2}  \left( \frac{5}{3}-\log \frac{-s}{M_V^2}
\right)  \right\}
\label{VFFRChTexpansion}\\
&& 
- \displaystyle\frac{4\,s^2}{F^4} \left\{  -\frac{F^4}{4M_V^4}
+ \bar\xi^{(4)}
-\frac{\Gamma_{88}^{(L)}-\Gamma_{90}^{(L)} }{32\pi^2}
\left( \frac{5}{3}-\log \frac{-s}{M_V^2} \right) \right\} +
\mathcal{O}\! \left(s^3\right)\, ,
\nonumber
\end{eqnarray}
where the $\bar\xi^{(2n)}$
are the relevant $\mathcal{O}(s^n)$ coefficients of the
low-energy expansion of  $\mF_{_{\rm NLO}}(s)$,
once the structure coming from the $\chi$PT one-loop diagram
has been subtracted from the $\pi\pi$ channel.
The separated contributions $\bar\xi^{(2n)}_{m_1,m_2}$
from each absorptive two-meson cut
$\mF_{_{NLO}}(s)|_{m_1,m_2}$  are provided in appendix~\ref{app.xi},
being each of them independent of the renormalization scale $\mu$.

By comparing the $\chi$PT expression~(\ref{VFFChPT}) to the \rcht\ low-energy
expansion~(\ref{VFFRChTexpansion}),
it is straightforward to estimate the chiral LECs
$L_9(\mu)$ and $C_{88}(\mu)-C_{90}(\mu)$
up to NLO in $1/N_C$:
\begin{equation}
\begin{array}{rl}
\qquad \qquad \qquad L_{9}(\mu) \,\, = &
\,\, \displaystyle\frac{F^2}{2M_V^2}\, \,
+\,\,  \bar\xi^{(2)} \, \,
+ \,\, \frac{\Gamma_9}{32\pi^2}\,\ln\frac{M_V^2}{\mu^2}  \,,
\\
C_{88}(\mu)\!-\!C_{90}(\mu)\,\,  =  &
\,\, -\displaystyle\frac{F^4}{4M_V^2}\,\,
+\,\,  \bar\xi^{(4)}\, \,
- \,\,  \frac{\Gamma_{88}^{(L)}-\Gamma_{90}^{(L)}}{32 \pi^2 }\!
\, \ln\Frac{M_V^2}{\mu^2} \,   ,
\end{array}\label{L9NLO}
\end{equation}
where
\begin{equation}
\Gamma_{88}^{(L)}-\Gamma_{90}^{(L)}\;\;
=\;\;\frac{3 G_V^2}{8 M_V^2} -
\frac{c_d^2}{4 M_S^2}+\frac{F_VG_V}{8 M_V^2}\;\;
=\;\;\frac{F^2-3 G_V^2}{8 M_S^2}- \frac{F^2+3 G_V^2}{8M_V^2}
\label{Gamma}
\end{equation}
matches the corresponding $\cO(p^6)$ running at NLO in $1/N_C$.
Note that the large--$N_C$ relations
$L_2=2L_1=\frac{G_V^2}{4 M_V^2}$,
$L_3=-\frac{3 G_V^2}{4M_V^2}+\frac{c_d^2}{2M_S^2}$
and $L_9=\frac{F_V G_V}{2 M_V^2}$~\cite{RChTa} have been used in eq.(\ref{matching}). The high-energy constraints
$F_V G_V=F^2$ and $2c_d^2=F^2- 3G_V^2$ of
eq.~(\ref{constraintpipi}) have been employed to obtain the result on the r.h.s. of eq.~(\ref{Gamma}).


 \subsection{Phenomenology}

 Using $M_V\simeq 0.77\,$GeV and $F\simeq 89\,$MeV, one gets the large-$N_C$ estimates from eq.~(\ref{L10LO}): $L_9\simeq 6.7 \cdot 10^{-3} $ and $C_{88}-C_{90} \simeq -4.5 \cdot 10^{-5}$.  At $\mu_0=770$~MeV,  the phenomenological determinations $L_9(\mu_0)=\left( 6.9\pm 0.7 \right) \cdot 10^{-3}$~\cite{ChPTp4,polychromatic} and $L_9(\mu_0)=\left(5.93 \pm 0.43\right)\cdot 10^{-3}, C_{88} (\mu_0)-C_{90} (\mu_0) =\left( -5.5 \pm 0.5 \right) \cdot 10^{-5}$~\cite{VFF_ChPT}, obtained respectively from an $\mathcal{O}(p^4)$ and an $\mathcal{O}(p^6)$ ChPT fit, agree approximately with the LO estimates.

Large--$N_C$ estimates are naively expected to approximate well the couplings at scales of the order of the relevant dynamics involved ($\mu \sim M_R$). However, they always carry an implicit error because of the uncertainty on $\mu$. This theoretical uncertainty is rather important in couplings generated through scalar meson exchange,
such as $L_8(\mu)$. In the present case, it also has a moderate importance. The size of the NLO corrections in $1/N_C$ to $L_{9}(\mu)$ and $C_{88}(\mu)-C_{90}(\mu)$
can be estimated by regarding their variations with $\mu$. These are respectively given by
\begin{equation}
\frac{\partial \, L_{9}(\mu)}{\partial \log \mu^2}
\,=\, -\, \frac{\Gamma_{9}}{32\pi^2}= -0.8\cdot 10^{-3} \, , \quad
\frac{\partial \, \left( C_{88}(\mu) - C_{90}(\mu) \right) }{\partial \log \mu^2}
= \frac{\Gamma_{88}^{(L)}-\Gamma_{90}^{(L)}}{32\pi^2}
\simeq 0.9\cdot 10^{-5} \, .
\label{eq.largeNCerror}
\end{equation}

So far, we have been working within a $U(3)_L\otimes U(3)_R$ framework, but we are actually interested on the couplings of the standard $SU(3)_L\otimes SU(3)_R$ chiral theory. Thus, a matching between the two versions of \chpt\ must be performed. Nonetheless, on the contrary to what happens with other matrix elements
({\it e.g.} the $S-P$ correlator~\cite{L8-nlo}), the spin--1 two-point functions do not gain contributions from the
$U(3)$--singlet chiral pseudo-Goldstone; the $\eta_1$ does neither enter at tree-level nor in the one-loop correlators. Therefore, the corresponding LECs are identical in both theories at leading and next-to-leading order in $1/N_C$:  $L_{9}(\mu)^{U(3)} = L_{9}(\mu)^{SU(3)}$,
$\left( C_{88}(\mu)-C_{90}(\mu) \right)^{U(3)}=\left( C_{88}(\mu)-C_{90}(\mu) \right)^{SU(3)}$.

The needed input parameters are defined in the chiral limit.
We take the ranges~\cite{ChPTp4,PDG} $M_V=\left( 770 \pm 5 \right)\,$MeV,
$M_S=\left(1090 \pm 110 \right)\,$MeV
and $F=\left(89\pm2 \right)\,$MeV.
The resonance couplings $G_V$ and $F_A$ can be fixed
in terms of $F$ and masses  if one considers the short-distance conditions
obeyed by the left--right correlator~\cite{polychromatic}.
The constraint
of eq.~(\ref{constraintLO}), coming from the vector form factor of the pion,
and those from the first and second Weinberg sum rules~\cite{SR} determine the
vector and axial-vector couplings
at LO in $1/N_C$~\cite{natxo-tesis,L10-nlo},
\begin{equation}
F_V^2=  F^2 \frac{M_A^2}{M_A^2-M_V^2}  , \qquad
G_V^2=F^2  \frac{M_A^2-M_V^2}{M_A^2} , \qquad
F_A^2=F^2 \frac{M_V^2}{M_A^2-M_V^2} , \label{WSRconstraints}
\end{equation}
with $M_A>M_V$. 
Due to the large width of the $a_1(1260)$ meson, the determination of the
Lagrangian parameter $M_A$ is far from trivial.
From the observed rates
$\Gamma (\rho^0 \to e^+ e^-)=(7.02\pm 0.13)$~keV~\cite{PDG}
and $\Gamma\left( a_1 \to \pi \gamma \right)= (650 \pm 250 )\,$keV~\cite{PDG}, and considering (\ref{WSRconstraints}), one finds $M_A=(938\pm13)$~MeV and $M_A=(960\pm80)$~MeV. Another large--$N_C$ determination of $M_A$ was obtained in ref.~\cite{jorge_vicent} from the study of the $\pi \to e \nu_e \gamma$ decay, which yields $M_A=(998\pm 49)$~MeV. We cannot use the information coming from $\Gamma (\rho \to 2\pi) = (149.4 \pm 1.0)$~MeV~\cite{PDG} in order to determine $M_A$,
since $G_V$ is constrained by eq.~(\ref{constraintpipi}) to be smaller
than $F/\sqrt{3}$, which results in $M_A< 940\,$MeV. In spite of the
dispersion of values for $M_A$, one gets a consistent description in the range
$M_A= (920 \pm 20 )\,$MeV, which we will take as our input.
The resulting numerical predictions for the LECs are
\begin{eqnarray}
L_9 (\mu_0 ) &=& \left( 7.9 \pm 0.4 \right) \cdot 10^{-3} \, , \nonumber \\
 C_{88} (\mu_0) - C_{90} (\mu_0)  &=& \left( -4.6 \pm 0.4 \right) \cdot 10^{-5} \,, \label{pheno2}
\end{eqnarray}
being $\mu_0$ the usual renormalization scale, $\mu_0=770\,$MeV.

\begin{table}[tb]
\begin{center}
\begin{tabular}{|c|r@{$.$}l | r@{$.$}l |}
\hline
  &  \multicolumn{2}{c|}{1st Approach} & \multicolumn{2}{c|}{2nd Approach}    \\[5pt]
  \hline \hline
  $10^3 \cdot L_9$ at LO &  $6$ & $68$ & $6$&$68$ \\ \hline
  $10^3 \cdot \bar{\xi}^{(2)}_{\pi\pi} $  & $0$&$11$  & $-0$&$04$ \\
 $10^3 \cdot \bar\xi^{(2)}_{P\pi} $ & $0$&$00$ &$0$&$00$ \\
  $10^3 \cdot \bar\xi^{(2)}_{A\pi} $ & $1$&$12$ & $1$&$00$  \\[5pt]
  \hline \hline
  $10^5 \cdot \left( C_{88}-C_{90} \right)$ at LO &  $-4$&$46$ & $-4$&$46$ \\ \hline
  $10^5 \cdot \bar\xi^{(4)}_{\pi\pi} $ & $0$&$76$   & $0$&$71$   \\
  $10^5 \cdot \bar\xi^{(4)}_{P\pi} $ &$0$&$00$ &$0$&$00$ \\
  $10^5 \cdot \bar\xi^{(4)}_{A\pi} $  & $-0$&$88$ & $-0$&$73$\\
\hline
\end{tabular}
\end{center}
\caption{Different contributions to the chiral couplings within the two numerical approaches explained in the text.}
\label{tab:tab1}
\end{table}

Alternatively, one could also use the phenomenological values for
$G_V$, $F_A$ and the axial-vector mass, instead of fixing them through the
Weinberg sum-rules. Thus, one may employ
$M_A= \left(1200 \pm 200 \right)\,$MeV~\cite{PDG},
and  $F_A=(120\pm 20 )\,$MeV, from the observed rate
$\Gamma\left( a_1 \to \pi \gamma \right)= (650 \pm 250 )\,$keV~\cite{PDG}.
The constraint of eq.~(\ref{constraintpipi})
implies that $G_V < F/\sqrt{3}$, so that we take the range
$G_V \in [40,50]\,$MeV.
For the remaining inputs $M_V$, $M_S$ and $F$, we consider the same values
used before, yielding the predictions
\begin{eqnarray}
L_9 (\mu_0 ) &=& \left( 7.6 \pm 0.6 \right) \cdot 10^{-3} \, , \nonumber \\
 C_{88} (\mu_0) - C_{90} (\mu_0)  &=& \left( -4.5 \pm 0.5 \right)
 \cdot 10^{-5} \,. \label{pheno1}
\end{eqnarray}

As it can be observed, the influence of using the first or the second
approach is not crucial at the present level of accuracy.
We take the values in (\ref{pheno2}), which include more theoretical
 constraints, as our final next-to-leading-order estimates for the LECs.

\begin{figure}
\begin{center}
\includegraphics[angle=0,width=7.45cm,clip]{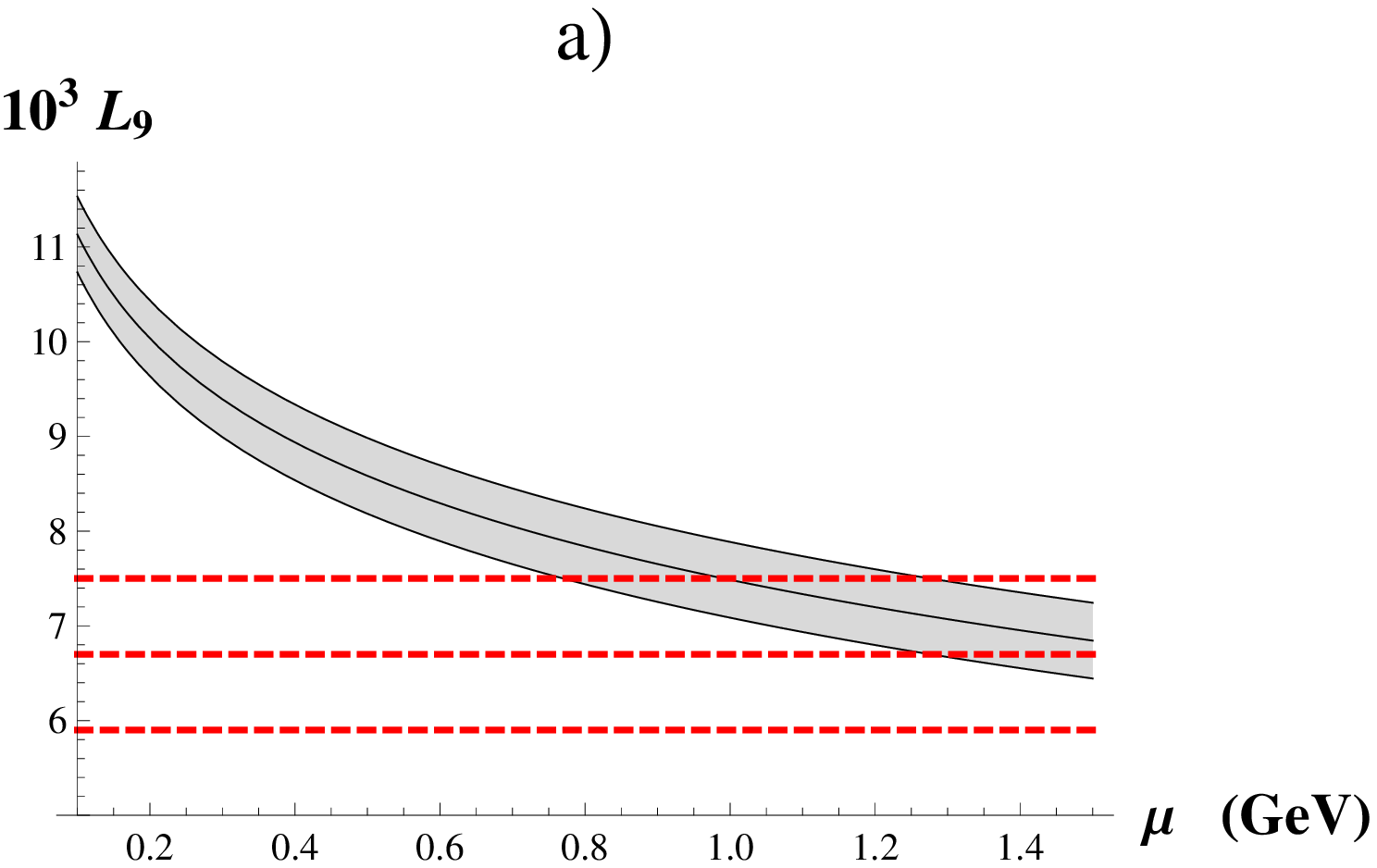}
\includegraphics[angle=0,width=7.45cm,clip]{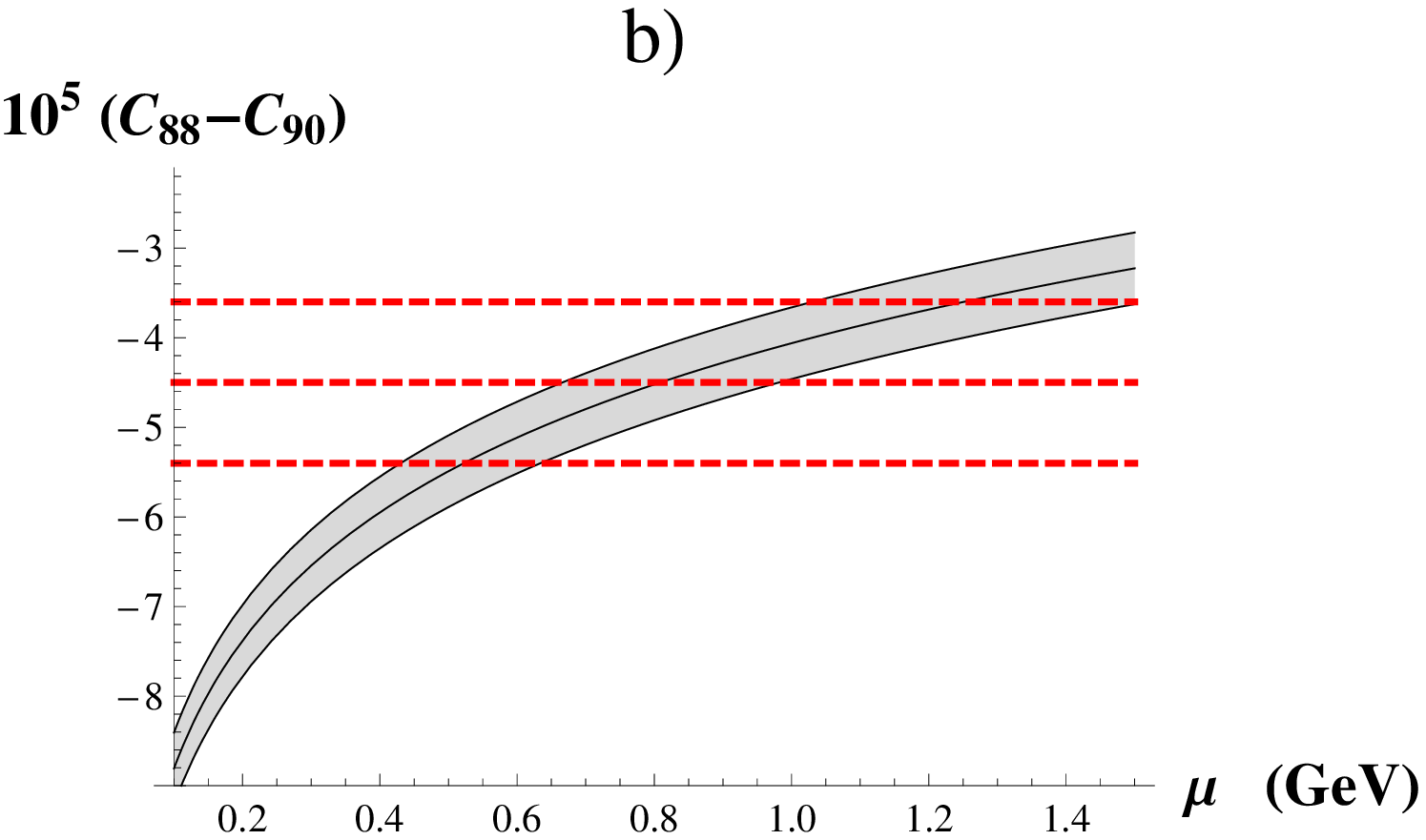}
\caption{{\small
The \rcht\ predictions (solid gray band)  for the \chpt\
$\cO(p^4)$ low-energy constant $L_9(\mu)$ (a)
and the $\cO(p^6)$ combination $C_{88}(\mu)-C_{90}(\mu)$  (b)
are compared to their large--$N_C$ estimates (red dashed) for
different values of the renormalization scale $\mu$.
The error of the large--$N_C$ estimate is given by
the naive saturation scale uncertainty from
eq.~(\protect\ref{eq.largeNCerror}).
}}
\label{fig.running}
\end{center}
\end{figure}

In table~\ref{tab:tab1} we present the different contributions to the LECs
within the first and second approaches.  A graphical comparison
of the NLO predictions and the large--$N_C$ estimates
has been made in figure~\ref{fig.running} for different values of the
renormalization scale $\mu$.

It is appropriate to note the appreciable increase of $L_9(\mu_0)$ from the large-$N_C$ prediction, for $\mu_0=770\,$MeV. In fact, the correction $\delta_{NLO}$ in eq.~(\ref{deltaNLO}) gets a contribution  from the $A\pi$ channel which is still comparable to that from the $\pi\pi$ one. This subleading contribution to $F_V G_V$, fixed through short-distance matching, increases the value of  $L_9$ by $1 \cdot 10^{-3}$, a quite sizeable shift. For details see appendix \ref{app.diagrams} and ref.~\cite{SanzCillero:2010zi}. 

\section{Conclusions} \label{sec:conclusions}

In this article we have completed the analysis of the  VFF
at NLO in $1/N_C$, initiated in ref.~\cite{RSP:05},
where the general framework was established.
We have considered  operators with more than one resonance
and have studied contributions from intermediate channels with resonances.
We get a well-behaved VFF at high-energies,
which goes to zero for $q^2\to \infty$~\cite{brodsky-lepage}.

Imposing that  each individual absorptive cut vanishes
at short distances,   one gets stringent constraints
on the structure of the VFF, which led to a prediction of the relevant
$\cO(p^4)$ and $\cO(p^6)$ \chpt\  couplings
up to NLO in $1/N_C$. The required inputs are the resonance
masses $M_V$, $M_A$ and $M_S$, 
and the pion decay constant $F$.  As expected for such a well-known
observable, the large--$N_C$ prediction provides already an excellent
estimate and the subleading corrections are relatively small.
At the reference scale $\mu_0=770$~MeV, we obtain
\begin{eqnarray}
L_9 (\mu_0 ) &=& \left(7.9\pm 0.4 \right) \cdot 10^{-3} \, , \nonumber \\
 C_{88} (\mu_0) - C_{90} (\mu_0)  &=&
 \left( -4.6 \pm 0.4\right) \cdot 10^{-5} \,.
\label{eq.final-det}
\end{eqnarray}
As the matching of \rcht\ with \chpt\ is complete up to NLO in $1/N_C$,
we fully  control   the running of the LECs up to that order
and, {\it e.g.}, we are able to predict $L_9(\mu)$
for any desired value of $\mu$.

\begin{table}
\begin{center}
\begin{tabular}{|c|c|c|}
\hline  & & \\[-5pt]
  &  $10^3 \cdot L_9 (\mu_0 ) $  &
  $10^5 \cdot   \left( C_{88} (\mu_0) - C_{90} (\mu_0) \right)$    \\[5pt]
\hline  & & \\[-5pt]
This work 1st & $7.9 \pm 0.4$ & $-4.6 \pm 0.4 $   \\
This work 2nd & $7.6 \pm 0.6 $ & $-4.5 \pm 0.5$   \\
Ref.~\cite{ChPTp4} & $6.9 \pm 0.7 $ & \\
Ref.~\cite{VFF_ChPT}  & $5.93 \pm 0.43 $ &  $-5.5 \pm 0.5 $ \\
Ref.~\cite{preVFF2} & $7.04 \pm 0.23 $ & \\
Ref.~\cite{GA} at $\cO(p^4)$& $6.54 \pm 0.15$ & \\ Ref.~\cite{GA} at $\cO(p^6)$
 & $5.50 \pm 0.40 $ & \\
 Ref.~\cite{Masjuan:2008fv}& $6.3 \pm 0.4$ &   \\
 [5pt]
 %
\hline
\end{tabular}
\end{center}
\caption{Comparison of our result with other determinations, being $\mu_0=770\,$MeV.}
\label{tab:tab2}
\end{table}


This result is in reasonable agreement with previous calculations~\cite{ChPTp4,VFF_ChPT,preVFF2,GA,Masjuan:2008fv}, see table~\ref{tab:tab2}, and shows once more the efficacy of \rcht\ to describe low-energy QCD matrix elements, specially if they are dominated by resonances.
It is important to remark not only that the amplitude is dominated by tree-level exchanges but also the fact that the one-loop corrections are not large.

Our determination of  $L_9 (\mu_0 )$ has a larger central value than the result obtained from an $\cO (p^6)$ chiral fit to the VFF~\cite{VFF_ChPT} at low energies, and it is closer to the chiral fit determination at $\cO (p^4)$~\cite{ChPTp4}. On the other hand, the ALEPH $\tau$-data analysis performed in~\cite{preVFF2}, which is also of $\cO (p^6)$ but takes higher-energy data into account, yields a value of the order of $7\cdot 10^{-3}$, much closer to our estimate.

In future works, we plan to study the pion scalar form-factor and the LECs $L_4(\mu)$ and $L_5(\mu)$, where the situation
is much less clear since, in that case, one has  contributions from broad resonance states like the $f_0(600)$.

\section*{Acknowledgments}

We wish to thank Jaroslav Trnka for collaboration at an early stage of this project. This work has been supported by the Universidad CEU Cardenal Herrera, by the Generalitat Valenciana (Prometeo/2008/069), by the Generalitat de Catalunya (SGR 2005-00916), by the Spanish Government (FPA2007-60323, FPA2008-01430, the Juan de la Cierva program and Consolider-Ingenio 2010 CSD2007-00042, CPAN), and by the European Union (MRTN-CT-2006-035482, FLAVIAnet). J.J.S.C. wants to thank IFAE, where part of this work has been done. 


\appendix


\section{Dispersion relations and loop contribution}
\label{app.disp-rel}

One may use a once--subtracted dispersion relation,
derived from the identity
\begin{eqnarray}
\Frac{\mF(s)}{s} & = & \Frac{1}{2\pi i} \, \oint \, \
\mathrm{d}t\; \Frac{\mF(t)}{t\,(t -  s) }  \, ,
\end{eqnarray}
where the integration is performed in the usual complex
circuit~\cite{L10-nlo}. The form-factor in the integrand can be written as
\begin{eqnarray}
\Frac{\mF(t)}{t}&=& \Frac{D(t)}{\left(M_V^2  -  t\right)^2} \, ,
\end{eqnarray}
where  $D(t)$ is an analytical  function except for the
unitarity logarithmic branch cut and the  single pole of $\frac{\mF(t)}{t}$
at $t=0$.
One gets then
\begin{eqnarray}
\label{eq.masterB}
\Frac{1}{s}\,\,\mF(s)&=& \Frac{1}{s}\,+\,  
\Frac{1}{s} \,\mF^{1\ell}(s)               
-  \Frac{\mbox{Re}D'(M_V^2)}{M_V^2-s}  +
\Frac{\mbox{Re}D(M_V^2)}{\left(M_V^2-s\right)^2} \, ,
\end{eqnarray}
where the  $\frac{1}{s}$ term on the r.h.s. is given by the integration
$\frac{1}{2\pi i}\int_{\theta=2\pi^-}^{\theta=0^+}\frac{dt}{t}\,\frac{\mF(t)}{
(t-s)}$, with ${  t=\epsilon\, e^{i\theta}  }$,
around $t=0$
of the function $\frac{\mF(t)}{t}\approx \frac{1}{t}+\cO(t^0)$,
and the different contributions
of each two-meson absorptive cut are given by the dispersive
integral,
\begin{eqnarray}
  \mF^{1\ell}(s) |_{m_1,m_2} &= & \lim_{\epsilon\to 0} \left[
\Frac{s}{\pi}\int_0^{M_V^2-\epsilon}
\!\!\! \mathrm{d}t\; \Frac{\mbox{Im}\mF(t)|_{m_1,m_2}}{
t\,( t\, -\, s)}
\, + \,
\Frac{s}{\pi}\int_{M_V^2+\epsilon}^\infty \!\!\! \mathrm{d}t\;
 \Frac{\mbox{Im}\mF(t)|_{m_1,m_2}}{t\,(t\, -\, s)}
\right. \nonumber
\\
&&\qquad \left. \,-\, \Frac{2 s}{\pi\epsilon}
\, \lim_{t\to M_V^2}
\left\{(M_V^2-t)^2\, \Frac{\mbox{Im}\mF(t)|_{m_1,m_2}}{t\,(t\,-\,s)}
\right\}\,\,
\right]\, .
\label{eq.F1loop}
\end{eqnarray}
Notice that if the threshold of the channel is above the resonance mass $M_V$,
then this expression gets simplified into the form
\begin{eqnarray}
\label{DeltaPiC}
  \mF^{1\ell}(s) |_{m_1,m_2} &= & \lim_{\epsilon\to 0}
\Frac{s}{\pi}\int_{(M_1+M_2)^2}^\infty \!\!\! \mathrm{d}t\;
 \Frac{\mbox{Im}\mF(t)|_{m_1,m_2}}{t\,(t\, -\, s)}
 \, ,
\end{eqnarray}
with $M_1$ ($M_2$) the mass of the $m_1$ ($m_2$) meson.

If we choose the on-shell mass scheme, without double poles in the perturbative expansion,
we have then
\begin{eqnarray}\label{eq.masterB}
\mF(t)&=& 1\,+\, \sum_{m_1,m_2} \mF^{1\ell}(t)|_{m_1,m_2}
-  \Frac{s\,\mbox{Re}D'(M_V^2)}{M_V^2-t}\, ,
\end{eqnarray}
where $\mbox{Re}D'(M_V^2)$ can be identified with
$-\frac{F_V^rG_V^r}{F^2}$ for a convenient renormalization scheme
of this combination of vector couplings~\cite{L8-nlo,L10-nlo,L8-Trnka}
(see appendix~\ref{app.diagrams} for further details).



\section{The spectral functions $\mathrm{Im} \, \mathcal{F}(s)|_{m_1,m_2}$} \label{app.ImF}

In this appendix we show the explicit form of the the spectral
functions of the different
two-particle absorptive cuts. First we present the functions obtained directly from the Feynman diagrams before imposing any short-distance constraint, {\it i.e.}, they are
badly behaved at high energies.
\begin{eqnarray}
\mathrm{Im} \, \mathcal{F}(s)|_{\pi\pi}&&=
\frac{F^2 \left(M_V^2-s\right)+s F_V G_V}{64 \pi  F^6 s^2 \left(s-M_V^2\right)}  \Bigg\{ 2 c_d^2 \left( M_S^4 \log \!\bigg(
  1+ \frac{s}{M_S^2}\bigg)  \left( -12 M_S^2 -6 s  \right)  +s^3 \right.\nonumber \\
   && +12 s M_S^4 \!\Bigg)\!+\!G_V^2
   \!\Bigg(\!s^3\!-\!6 M_V^2 \left(M_V^2\!+\!2 s\right)    \left. \! \left(\! \log\! \bigg(1\!+\!\frac{s}{M_V^2}\bigg) \left(  2 M_V^2 \!+\!s  \right)\!-\!2s\!\right)\!\right)\! \Bigg\} \nonumber  \\
   && +  \frac{ s^2 G_V\left(F^2
   \left(F_V+2 G_V\right) \left(M_V^2-s\right)+2 s F_V G_V^2 \right)}{64 \pi  F^6 \left(s-M_V^2\right)^2}+\frac{s}{64 \pi  F^2} \,, \label{Impipi1} \\
   && \nonumber \\
   \mathrm{Im} \, \mathcal{F}(s)|_{P\pi}&&=
   \frac{\sqrt{2} c_d F_V  \lambda_1^{\mathrm{SP}} \lambda _1^{\mathrm{PV}}} {32 \pi  F^4s \left(s-M_V^2\right)}\Bigg\{
   3 M_P^4 \left(4 M_S^2+s\right)-3 M_P^2 \left(2 M_S^2+s\right)^2 -M_P^6\nonumber \\
   && -6 M_S^2 \!\left(M_S^2-M_P^2\right)\! \left(-M_P^2+2
   M_S^2+s\right)\! \log \!\bigg(1\!+\!\frac{s-M_P^2}{M_S^2}\bigg)+12 s M_S^4+s^3 \Bigg\} \nonumber \\
   && -
    \frac{F_V G_V {\lambda _1^{\mathrm{PV}}}^2} {32 \pi  F^4 s \left(s\!-\!M_V^2\right)} \Bigg\{
   3 M_P^2\! \left(12 s M_V^2\!+\!4 M_V^4\!+\!s^2\right)\!+6 M_V^2 \!\left(-3 M_P^2
  \! \left(M_V^2\!+\!s\right) \right.\nonumber \\
   &&\left. +M_P^4+5 s M_V^2+2 M_V^4+2 s^2\right) \log \!\bigg(1\!+\!\frac{s-M_P^2}{M_V^2}\bigg)-3 M_P^4 \left(4
   M_V^2+s\right) \nonumber \\
   && +M_P^6-s \left(24 s M_V^2+12 M_V^4+s^2\right)  -    \frac{2 s\left(s-M_P^2\right)^3}{s-M_V^2}\Bigg\}  \,, \label{ImPpi1} \\
   \nonumber \\
    \mathrm{Im} \, \mathcal{F}(s)|_{A\pi}&&=
    \frac{-G_V\left(s\!-\!M_A^2\right)^2}{32 F^4 \pi M_A^2 s \left(s\!-\!M_V^2\right)^2} \Bigg\{ \!F_A\! \left((2 \kappa \!+\!\sigma ) M_A^4\!+\!4 s (\kappa \!+\!\sigma ) M_A^2\!+\!s^2 \sigma \right) \!\left(s\!-\!M_V^2\right)\nonumber \\ && -F_V \left(s-M_A^2\right) \left((2 \kappa +\sigma )^2 M_A^4+2 s \left(\kappa ^2+4 \sigma  \kappa +2
   \sigma ^2\right) M_A^2+s^2 \sigma ^2\right)  \Bigg\} \phantom{\frac{1}{2}}\!\!\!\!\nonumber \\ &&
    -\frac{G_V }{32 F^4 \pi M_A^2 s \left(s-M_V^2\right)}\Bigg\{ 6 \log \!\bigg(1+\frac{s-M_A^2}{M_V^2}\bigg) \left(F_A \left(s-M_V^2\right) \left(M_A^2-M_V^2\right) \right. \nonumber \\ && \left.
   \left(\kappa  M_A^2\!+\!\sigma  \!\left(M_V^2\!+\!s\right)\!\right)\!+\!F_V \!\left(\!\left(M_A^2\!-\!s\right) \!\left(M_A^2\!-\!M_V^2\right)\!
   \left(M_V^2\!+\!s\right) \sigma ^2\!+\!2 \kappa  M_A^2 \right. \right.\phantom{\frac{1}{2}} \!\!\!\!\nonumber \\ && \left. \left.
    \left(\!M_A^2\!-\!s\!\right)\! \left(\!M_A^2\!-\!M_V^2\!\right)\! \sigma \!+\!\kappa ^2 M_A^2 \!
   \left(\!3 M_A^4\!-\!5 \!\left(\!M_V^2\!+\!s\!\right) \!M_A^2\!+\!\left(\!M_V^2\!+\!2 s\!\right)\! \left(\!2 M_V^2\!+\!s\!\right)\!\right)\!\right)\!\right)\!\phantom{\frac{1}{2}}\!\!\!\!
     \nonumber \\ &&
   M_V^2 \!+\!\left(\!M_A^2\!-\!s\!\right)\! \left(\!F_A\! \left(\!s\!-\!M_V^2\!\right)\! \left((3 \kappa \!+\!\sigma ) M_A^4\!+\!\left((3 \sigma\! \!-6 \kappa )
   M_V^2\!+\!s (3 \kappa \!+\!4 \sigma \!)\right)\! M_A^2\right. \right. \phantom{\frac{1}{2}}\!\!\!\!\nonumber \\ && \left. \left.
   +\sigma \! \left(s^2\!-\!6 M_V^4\!-\!3 s M_V^2\right)\!\right)\!+\!F_V
   \!\left(\!\left(M_A^2\!-\!s\right) \!\left(M_A^4\!+\!4 s M_A^2\!-\!6 M_V^4\!+\!s^2 \right. \right. \right.\phantom{\frac{1}{2}}\!\!\!\! \nonumber \\ && \left. \left. \left.
   +3 \left(M_A^2-s\right) M_V^2\right) \sigma ^2+6 \kappa
   M_A^2 \left(M_A^2-s\right) \left(M_A^2-2 M_V^2+s\right) \sigma  \right. \right. \phantom{\frac{1}{2}}\!\!\!\!\nonumber \\ && \left. \left.
   +\kappa ^2 M_A^2 \left(7 M_A^4-8 \left(3 M_V^2+s\right)
   M_A^2+12 M_V^4+s^2+24 s M_V^2\right)\right)\right)
  \Bigg\} \phantom{\frac{1}{2}}\!\!\!\!\nonumber \\ &&
    +\frac{\sqrt{2} c_d\lambda_1^{\mathrm{SA}}}{32 F^4 \pi  s \left(s-M_V^2\right)}  \Bigg\{    6 \log \!
   \bigg(1+\frac{s-M_A^2}{M_S^2}\bigg) \left(F_V \left(2 \kappa  M_S^4+(\kappa -\sigma )M_S^2\right. \right. \nonumber \\ && \left. \left.
    \left(s-M_A^2\right)
   +(\kappa +\sigma ) M_A^2 \left(s-M_A^2\right)\right)+F_A \left(M_A^2-M_S^2\right) \left(M_V^2-s\right)\right)
   M_S^2 \phantom{\frac{1}{2}}\!\!\!\!\nonumber \\ &&
   +\left(M_A^2-s\right) \left(F_V \left(3 \sigma  \left(s-M_A^2\right) \left(M_A^2-2 M_S^2+s\right)  \right. \right.\phantom{\frac{1}{2}}\!\!\!\!\nonumber \\ && \left. \left.
   +\kappa
   \left(4 s M_A^2\!-\!5 M_A^4\!+\!12 M_S^4\!+\!s^2\right)\!\right)\!+\!3 F_A\! \left(M_A^2\!-\!2 M_S^2\!+\!s\right)\!
   \left(M_V^2\!-\!s\right)\!\right)\!\! \Bigg\}\,, \label{ImApi1} 
       \end{eqnarray}
where we have used the combination of couplings $\kappa$ and $\sigma$,
\begin{equation}
\kappa\,=\, -2 \lambda_2^{\mathrm{VA}} + \lambda_3^{\mathrm{VA}} \,,  \qquad \qquad
\sigma \,=\,   2 \lambda_2^{\mathrm{VA}} -2 \lambda_3^{\mathrm{VA}} + \lambda_4^{\mathrm{VA}} + 2 \lambda_5^{\mathrm{VA}} \,.
\end{equation}

After considering the constraints explained in section~3,
\begin{equation}
\begin{array}{rlrl}
F_V G_V&= F^2 \,, \qquad \quad \phantom{\displaystyle\frac{1}{2}} &   3\,G_V^2+2\,c_d^2 &= F^2 \,,  \nonumber \\
\lambda_1^{\mathrm{PV}}&=0\,, \qquad \quad \phantom{\displaystyle\frac{1}{2}}& \kappa&=0  \,, \nonumber \\
\kappa + \sigma &= \displaystyle\frac{F_A}{F_V} \,, \qquad \quad & \lambda_1^{\mathrm{SA}}&= -\displaystyle\frac{F_A \,G_V\left(M_A^2 - 4\,M_V^2\right)}{3\sqrt{2} M_A^2 c_d\,F_V} \,,
\end{array}
\label{constraintsbis}
\end{equation}
the imaginary part of each absorptive cut vanishes at short-distances and
the following expressions are found,
\begin{eqnarray}
\mathrm{Im} \, \mathcal{F}(s)|_{\pi\pi}&&=    \frac{M_V^2}{32 \pi  F^4 s^2 \left(s\!-\!M_V^2\right)^2}\Bigg\{
 3M_S^4 \!\left(F^2\!-\!3 G_V^2\right)\! \left(M_V^2\!-\!s\right) \!\log \!\bigg(1\!+\!\frac{s}{M_S^2}\bigg)\! \left(2 M_S^2 + s \right) \nonumber \\ &&
 +G_V^2 M_V^2 \Bigg( \log\! \bigg(1+\frac{s}{M_V^2}\bigg) \left(  -6s^3 -9s^2 M_V^2+6M_V^6+9sM_V^4\right) +13 s^3    \nonumber \\ &&
 \Bigg. -6 s^2 M_V^2 -6 s M_V^4 \Bigg)
 +6 s M_S^4
   \left(F^2-3 G_V^2\right) \left(s-M_V^2\right)    \Bigg\} \,,\label{Impipi} \\ \nonumber \\
 \mathrm{Im} \, \mathcal{F}(s)|_{P\pi}&&=0 \, ,\label{ImPpi}\\ \nonumber \\
  \mathrm{Im} \, \mathcal{F}(s)|_{A\pi}&&= \frac{F_A^2 G_V^2 \left(M_V^2\!-\!M_A^2\right)}{32 \pi  F^6 s M_A^2 \left(s\!-\!M_V^2\right)^2}\Bigg\{
   M_A^4 \Bigg(2 M_S^2 \left(M_V^2\!-s\right) \left(\log\!
   \bigg(1\!+\frac{s-M_A^2}{M_S^2}\bigg)\!-\!1\right)  \nonumber \\ &&
   +4 s M_V^2-7 M_V^4-3 s^2\Bigg)+2 M_A^2 \left(s^2 M_V^2 \left(3 \log\!
   \bigg(1+\frac{s-M_A^2}{M_V^2}\bigg)-2\right) \right. \nonumber \\ && \left.
   +M_S^4 \left(s-M_V^2\right) \log\!
   \bigg(1+\frac{s-M_A^2}{M_S^2}\bigg)-M_S^2 \left(s-M_V^2\right) \right. \nonumber \\ && \left.
    \left(s\!-\!4 M_V^2 \left(\log\!
   \bigg(1\!+\! \frac{s\!-\!M_A^2}{M_S^2}\bigg)\!-\!1\right)\right)\!-\!3 M_V^6 \left(\log\!
   \bigg(1\!+\!\frac{s\!-\!M_A^2}{M_V^2}\bigg)\!-1\right)\right)\nonumber \\ &&
   +M_V^2\! \left(s^2 M_V^2 \!\left(7\!-\!6 \log\!
   \bigg(1\!+\!\frac{s\!-\!M_A^2}{M_V^2}\bigg)\right)\!+\!8 M_S^4 \left(M_V^2\!-\!s\right) \log\!
   \bigg(1\!+\!\frac{s\!-\!M_A^2}{M_S^2}\bigg) \right. \nonumber \\ && \left.
   +6 M_V^6 \log\! \bigg(1\!+\!\frac{s\!-\!M_A^2}{M_V^2}\bigg)\!+\!8 s M_S^2
   \left(s\!-\!M_V^2\right)\!-\!6 s M_V^4\!\right)\!+\!2 s M_A^6\!+\!M_A^8\!
   \Bigg\} .\label{ImApi}
\end{eqnarray}



\section{Next-to-leading-order corrections $\mathcal{F}_{_{\rm NLO}}(s)|_{m_1,m_2}$}
\label{app.F1loop}

In this appendix we show the explicit form of the NLO corrections
generated by the considered two-particle absorptive cuts,
eqs. (\ref{Impipi}), (\ref{ImPpi}) and (\ref{ImApi}),
which have been calculated by using the dispersive method discussed
in appendix \ref{app.disp-rel}. Below, we have summed up the $\delta_{_{\rm NLO}}$ contribution
to $\mF^{1\ell}(s)$, as seen in eq.~(\ref{deltaNLO}), being the different
$\mathcal{F}_{_{\rm NLO}}(s)|_{m_1m_2}$ well-behaved at high energies:
\begin{eqnarray}
\mathcal{F}_{_{\rm NLO}}&&\!\!\!\!(s)|_{\pi\pi}= \frac{M_V^2}{64 \pi ^2 F^4 s \left(s-M_V^2\right)^2} \Bigg\{
-12 M_S^6 \left(F^2-3 G_V^2\right) \left(s-M_V^2\right)
   f \! \left(s,M_S^2\right)  \Bigg. \nonumber \\
   && \Bigg. -6 M_S^4 \left(F^2-3 G_V^2\right) \left(s-M_V^2\right)
  \left(s f\!\left(s,M_S^2\right)+2 \log \bigg(\frac{-s}{M_S^2}\bigg) -2 \right)\nonumber  \\
   && \Bigg. +G_V^2 M_V^2 \Bigg(  -6 \left (3s^2 M_V^2 -3s M_V^4
-2M_V^6+2s^3 \right)  f \!\left( s,M_V^2\right)  \nonumber \\
&&\Bigg.\left.
 + s^2 \left(  -26 \log \bigg(\frac{-s}{M_V^2}\bigg)+27\right) +12
   M_V^4 \left( \log
   \bigg(\frac{-s}{M_V^2}\bigg)-1\right) \right. \Bigg.\nonumber  \\
   && \left. \Bigg.  +3 s M_V^2 \left(4
   \log \bigg(\frac{-s}{M_V^2}\bigg)-5\right)\right) +3 s M_S^2 \left(F^2-3 G_V^2\right)
   \left(s-M_V^2\right)
 \Bigg\}  \,, \label{VFFNLOpipi} \\ \nonumber \\
 %
 \mathcal{F}_{_{\rm NLO}}&&\!\!\!\!(s)|_{P\pi}=0 \,,  \label{VFFNLOPpi}\\ \nonumber \\
 %
   \mathcal{F}_{_{\rm NLO}}&&\!\!\!\!(s)|_{A\pi}= -\frac{F_A^2 G_V^2 \left(M_A^2-M_V^2\right)}{32 \pi ^2 F^6 s M_A^2 M_V^4 \left(s-M_V^2\right)^2}
   \Bigg\{ M_A^4 M_V^4\Bigg( 2 s M_S^2 \left(M_V^2-s\right)
 g\left(  s,M_A^2,M_S^2\right)  \nonumber \\
   && \Bigg. \left.
   -6 s^2 \log \!\bigg(\!1-\frac{M_V^2}{M_A^2}\!\bigg)\!+\log \!\bigg(\!1-\frac{s}{M_A^2}\!\bigg)\!\left( 3 s^2
  \!+\!2 M_S^2\! \left(M_V^2-s\right)\!  +\!7 M_V^4 \!-\!4 s M_V^2\!\right) \!\!\right) \Bigg. \nonumber  \\
   && \Bigg. +s M_V^6 \left(M_V^2 \left(-6
 \left(s^2-M_V^4\right)  g\left(  s,M_A^2,M_V^2\right) +6 M_V^2 \left(\log
   \bigg(1-\frac{s}{M_A^2}\bigg)-1 \right. \right. \right. \Bigg. \nonumber \\
   && \Bigg. \left. \left. \left.
   +\log \bigg(\frac{M_A^2}{M_V^2}\bigg)\!\right)\!+s\! \left(\!-7 \log \!
   \bigg(1-\frac{s}{M_A^2}\bigg)\!-6 \log \!\bigg(\frac{M_A^2}{M_V^2}\bigg)\!+\log \!
   \bigg(1-\frac{M_V^2}{M_A^2}\bigg)\!+6\!\right)\!\right)\right. \Bigg. \nonumber \\
   && \Bigg.\left.
   +8 M_S^4 \!\left(M_V^2-s\right)\!
  g\!\left(  s,M_A^2,M_S^2\right) \! -\!8 M_S^2\! \left(s-M_V^2\right)\!\left(\!\log
   \!\bigg(1\!-\!\frac{s}{M_A^2}\!\bigg)\!+\log \!\bigg(\frac{M_A^2}{M_S^2}\bigg)\!\right.\right. \Bigg. \nonumber \\
   && -1\!\Bigg)\!\Bigg)\!
   +M_A^2 M_V^4 \!\left(\!M_V^2
\! \left(\!6 \left(s^3-s M_V^4\right)  g\left(  s,M_A^2,M_V^2\right)+s^2 \!\left(\!4 \log
   \bigg(1-\frac{s}{M_A^2}\bigg) \right. \right. \right. \Bigg. \nonumber \\
   && \Bigg. \left. \left. \left. +2 \log \bigg(1-\frac{M_V^2}{M_A^2}\bigg)-7\right)-6 M_V^4 \log
   \bigg(1-\frac{s}{M_A^2}\bigg)+7 s M_V^2\right) \right. \Bigg. \nonumber \\
   && \Bigg. \left. +2 s M_S^4 \left(s-M_V^2\right)
  g\left(  s,M_A^2,M_S^2\right)  \right. \Bigg. \nonumber \\
   && \Bigg. \left. +2 M_S^2 \!\left(s-M_V^2\right)\! \left(\!4 s M_V^2
  g\left(  s,M_A^2,M_S^2\right) \!+s \!\left(\!\log \!\bigg(1-\frac{s}{M_A^2}\bigg)\!+\log\!
   \bigg(\frac{M_A^2}{M_S^2}\bigg)\!-1\!\right) \right. \right. \Bigg. \nonumber \\
   && \Bigg. \left. \left. +4 M_V^2 \log \bigg(1-\frac{s}{M_A^2}\bigg)\right)\right)+M_A^8 \left(s^2
   \log \bigg(1-\frac{M_V^2}{M_A^2}\bigg)-M_V^4 \log \bigg(1-\frac{s}{M_A^2}\bigg)\right) \Bigg. \nonumber \\
   && \Bigg. +s M_A^6 M_V^2 \!\left(\!M_V^2
  \! \left(\!-2 \log \!\bigg(1-\frac{s}{M_A^2}\bigg)\!-\!1\!\right)\!+2 s \log
  \! \bigg(1-\frac{M_V^2}{M_A^2}\bigg)\!+s\right)\!\Bigg\} \,, \label{VFFNLOApi}
\end{eqnarray}
where the functions $f(s,M^2)$ and $g(s,M_1^2,M_2^2)$ have been introduced for simplicity,
\begin{eqnarray}
f\left( s, M^2\right) &=& \frac{1}{s} \left( \mathrm{Li}_2 \bigg( 1 + \frac{s}{M^2} \bigg) - \frac{\pi^2}{6} \right)  \,, \nonumber \\
g\left(s, M_1^2, M_2^2 \right) &=& \frac{1}{s} \left( \mathrm{Li}_2 \bigg( 1 + \frac{s}{M_2^2} - \frac{M_1^2}{M_2^2} \bigg) - \mathrm{Li}_2 \bigg( 1 - \frac{M_1^2}{M_2^2} \bigg) \right) \,.
\end{eqnarray}


\section{NLO contributions to $L_9(\mu)$ and $C_{88}(\mu)-C_{90}(\mu)$} \label{ap:C}
\label{app.xi}

In this appendix we give the full expressions of the NLO contributions
to $L_9(\mu)$ and $C_{88}(\mu)-C_{90}(\mu)$, following
the notation of eqs.~(\ref{VFFRChTexpansion}) and (\ref{L9NLO}),
{\it i.e.}, $\bar\xi^{(2)}_{m_1,m_2}$ and $\bar\xi^{(4)}_{m_1,m_2}$:
\begin{eqnarray}
 \bar\xi^{(2)}_{\pi\pi}&=&  \frac{1}{768 \pi^2 F^2}   \Bigg\{ F^2 \left(6 \log \!\bigg(\frac{M_S^2}{M_V^2}\bigg)-11\right)+G_V^2 \left(38-18 \log\!
   \bigg(\frac{M_S^2}{M_V^2}\bigg)\right) \Bigg\}   \,, \\ && \nonumber \\
  \bar\xi^{(2)}_{P\pi}&=& 0     \,, \\ && \nonumber \\
  \bar\xi^{(2)}_{A\pi}&=&   \frac{F_A^2 G_V^2}{128 \pi^2 F^4 M_A^2 M_V^8 \left(M_A^2-M_S^2\right)} \Bigg\{
   2 M_A^{10} \left(M_S^2-M_A^2\right) \log \!\bigg(1-\frac{M_V^2}{M_A^2}\bigg) \nonumber \\ &&
   -2 M_A^8 M_V^2
   \left(M_A^2-M_S^2\right) \left(\log \!\bigg(1-\frac{M_V^2}{M_A^2}\bigg)+1\right)+M_A^6 M_V^4
   \left(M_A^2-M_S^2\right)\nonumber \\ &&
    \left(\!16 \log \!\bigg(1\!-\!\frac{M_V^2}{M_A^2}\bigg)\!-\!3\right)\!+\!M_A^4 M_V^6\! \left(\!M_S^2 \!\left(\!-2
   \log\! \bigg(\frac{M_A^2}{M_S^2}\bigg)\!+\!16 \log\! \bigg(1\!-\!\frac{M_V^2}{M_A^2}\bigg)\!-11\right) \right. \nonumber \\ && \left.
   +M_A^2 \left(11\!-\!16 \log\!
   \bigg(1\!-\!\frac{M_V^2}{M_A^2}\bigg)\right)\right)\!+\!M_A^2 M_V^8 \left(M_S^2 \left(10 \log\!
   \bigg(\frac{M_A^2}{M_S^2}\bigg)\!-\!12 \log \!\bigg(\frac{M_A^2}{M_V^2}\bigg) \right. \right. \nonumber \\ && \left. \left.
   -2 \log\!
   \bigg(1-\frac{M_V^2}{M_A^2}\bigg)+11\right)+M_A^2 \left(12 \log\! \bigg(\frac{M_A^2}{M_V^2}\bigg)+2 \log\!
   \bigg(1-\frac{M_V^2}{M_A^2}\bigg)-11\right)\right) \nonumber \\ &&
   +M_V^{10} \left(M_A^2 \left(-6 \log\!
   \bigg(\frac{M_A^2}{M_V^2}\bigg)+2 \log \!\bigg(1-\frac{M_V^2}{M_A^2}\bigg)+5\right) \right. \nonumber \\ && \left.
    -M_S^2 \left(8 \log\!
   \bigg(\frac{M_A^2}{M_S^2}\bigg)-6 \log\! \bigg(\frac{M_A^2}{M_V^2}\bigg)+2 \log\!
   \bigg(1-\frac{M_V^2}{M_A^2}\bigg)+5\right)\right)
 \Bigg\}  \,, \\ \nonumber \\
 \bar\xi^{(4)}_{\pi\pi}&=&   \frac{1}{3072 \pi ^2  M_V^2} \Bigg\{
 2 F^2 \left(11-6 \log\! \bigg(\frac{M_S^2}{M_V^2}\bigg)\right)+G_V^2 \left(36 \log\!
   \bigg(\frac{M_S^2}{M_V^2}\bigg)-11\right)\Bigg\} \nonumber \\ &&
   + \frac{\left( F^2-3 G_V^2\right)}{3072 \pi ^2  M_S^2}
  \Bigg\{ 12 \log\!
   \bigg(\frac{M_S^2}{M_V^2}\bigg)-19
 \Bigg\}  \,, \\ &&\nonumber \\
  \bar\xi^{(4)}_{P\pi}&=& 0     \,, \\ && \nonumber \\
  \bar\xi^{(4)}_{A\pi}&=&   \frac{-F_A^2 G_V^2}{384 \pi ^2 F^2 M_A^2 M_V^{10} \left(M_A^2-M_S^2\right)^2 \left(M_A^2-M_V^2\right)} \Bigg\{
    -6 M_A^{10} M_V^2 \left(M_A^2-M_S^2\right)^2 \nonumber \\ &&
    +8 M_V^{14}\! \left(\!M_S^2 \!\left(\!\log\!
   \bigg(\frac{M_A^2}{M_S^2}\bigg)\!+\!1\!\right)\!-\!M_A^2\!\right)\!-\!M_V^{12}\! \left(\!2 M_A^2 M_S^2 \!\left(\!3 \log\!
   \bigg(\frac{M_A^2}{M_S^2}\bigg)\!+\!6 \log\! \bigg(\frac{M_A^2}{M_V^2}\bigg) \right.\right. \nonumber \\ && \left.\left.
   -6 \log\!
   \bigg(1\!-\!\frac{M_V^2}{M_A^2}\bigg)\!+\!4\!\right)\!+\!M_S^4 \!\left(\!12 \log\! \bigg(\frac{M_A^2}{M_S^2}\bigg)\!-\!6 \log\!
   \bigg(\frac{M_A^2}{M_V^2}\bigg)\!+\!6 \log\! \bigg(1\!-\!\frac{M_V^2}{M_A^2}\bigg)\!+\!5\!\right) \right. \nonumber \\ && \left.
   +M_A^4 \left(-6 \log\!
   \bigg(\frac{M_A^2}{M_V^2}\bigg)+6 \log\! \bigg(1-\frac{M_V^2}{M_A^2}\bigg)-13\right)\right)-3 M_A^2 M_V^{10}
   \Bigg(M_A^2 M_S^2 \nonumber \\ && \left.
    \left(\!5 \log\! \bigg(\frac{M_A^2}{M_S^2}\bigg)\!-\!12 \log\!
   \bigg(\frac{M_A^2}{M_V^2}\bigg)\!+\!22\!\right)\!+\!M_S^4\! \left(\!-9 \log\! \bigg(\frac{M_A^2}{M_S^2}\bigg)\!+\!6 \log\!
   \bigg(\frac{M_A^2}{M_V^2}\bigg)\!-\!13\right) \right. \nonumber \\ && \left.
   +M_A^4 \left(6 \log\! \bigg(\frac{M_A^2}{M_V^2}\bigg)-9\right)\right)+2
   M_A^4 M_V^8 \left(M_A^2 M_S^2 \left(8 \log\! \bigg(\frac{M_A^2}{M_S^2}\bigg)-18 \log\!
   \bigg(\frac{M_A^2}{M_V^2}\bigg) \right. \right. \nonumber \\ && \left. \left.
   -54 \log\! \bigg(1\!-\!\frac{M_V^2}{M_A^2}\bigg)\!+\!71\!\right)\!+\!9 M_S^4 \!\left(\!\log\! \bigg(\frac{M_A^2}{M_V^2}\bigg)\!\!-\log\!
   \bigg(\frac{M_A^2}{M_S^2}\bigg)\!+\!3 \log\!
   \bigg(1\!-\!\frac{M_V^2}{M_A^2}\bigg)\!-\!4\!\right) \nonumber \right.\\ && \left.
   +M_A^4 \left(9 \log\! \bigg(\frac{M_A^2}{M_V^2}\bigg)+27 \log\!
   \bigg(1-\frac{M_V^2}{M_A^2}\bigg)-35\right)\right)-M_A^6 M_V^6 \left(M_A^2-M_S^2\right) \nonumber \\ &&
    \left(\!M_S^2 \!\left(\!3 \log\!
   \bigg(\frac{M_A^2}{M_S^2}\bigg)\!-\!96 \log\! \bigg(1\!-\!\frac{M_V^2}{M_A^2}\bigg)\!+\!47\!\right)\!+\!M_A^2 \!\left(\!96 \log\!
   \bigg(1\!-\!\frac{M_V^2}{M_A^2}\bigg)\!-\!47\!\right)\!\right)\nonumber \\ &&
   +\!3 M_A^8 M_V^4\!\left(\!M_A^2\!-\!M_S^2\!\right)^2 \!\left(\!18 \log\!
   \bigg(1\!-\!\frac{M_V^2}{M_A^2}\bigg)\!-\!1\!\right)\!-\!6 M_A^{12} \!\left(\!M_A^2\!-\!M_S^2\!\right)^2\! \log\!
   \bigg(1\!-\!\frac{M_V^2}{M_A^2}\bigg)
  \!\Bigg\}\,.  \nonumber \\ &&
\end{eqnarray}


\section{Description in terms of Feynman diagrams}
\label{app.diagrams}

The subleading corrections can be calculated by means of dispersive relations. Once the NLO absorptive parts of $\mathcal{F}_{R \chi T} \left(s\right)$ are known, one can reconstruct the full form factor up to appropriate subtraction terms.
Alternatively, we can compute and separate  the tree-level and one-loop amplitudes in the form
\begin{eqnarray}
\mathcal{F}_{R \chi T} \left(s\right) &=&
1\,+\, \Frac{F_V  G_V }{F^2}
 \Frac{s}{M_V^{ 2}\, -\, s}
 \, +\,  \Frac{2\widetilde{L}_9 }{F^2} \, s \,\,
\,\, + \sum_{m_1,m_2}
\mathcal{F}(s) |_{m_1,m_2} \,, \label{VFFRChT}
\end{eqnarray}
where the one-loop diagrams $\mathcal{F}(s) |_{m_1,m_2}$
can be rewritten by means
of a once-subtracted dispersion relation in the form
\begin{equation}
\sum_{m_1,m_2}
\mathcal{F}(s) |_{m_1,m_2} =
\sum_{m_1,m_2}
\mathcal{F}^{1\ell}(s) |_{m_1,m_2}
 \, +\,  \Frac{2\hat{\delta}_2}{F^2} \, s \,\,
 \,\,+\,\, \hat{\delta}_0
 \Frac{s}{M_V^{  2}\, -\, s}
 \,\,+\,\, \hat{\delta}_{-2}
 \Frac{s}{(M_V^{ 2}\, -\, s)^2}
\, .
 \label{eq.loop-contr}
\end{equation}
The finite part of the loops is contained in
the once-subtracted dispersive functions $\mathcal{F}^{1\ell}(s) |_{m_1,m_2} $,
 fully determined by the imaginary
part of Im$\mathcal{F}(s) |_{m_1,m_2}$ through
eq.~(\ref{eq.F1loop}).
The real parameters $\hat{\delta}_{-2,0,2}$ contain the ultraviolet
divergences of the loops, being $\hat{\delta}_{0}$ and $\hat{\delta}_{-2}$
the real part of the pole residues.
The local \rcht\ coupling $\widetilde{L}_9$ renormalizes $\hat{\delta}_2$,
the combination $F_V G_V$ cancels the divergences in $\hat{\delta}_0$
and a convenient shift of the mass, $M_V^{(B)\,\,2}=M_V^{  2}
+\delta M_V^2$ removes the divergent part of $\hat{\delta}_{-2}$.
Indeed, we will work in the on-shell scheme and the counterterm
$\delta M_V^2$ will be chosen to  completely kill $\hat{\delta}_{-2}$.

In order to finish the short-distance matching we just need to take into account
that the once-subtracted loop contribution behaves at short distances like
\begin{equation}
\sum_{m_1,m_2}
\mathcal{F}^{1\ell}(s) |_{m_1,m_2}\,\,\, \stackrel{s\to\infty}{\longrightarrow}
\,\,\, \delta_0\, \,\,+\,\,\, \cO(s^{-1})\, ,
\end{equation}
with $\delta_0$ a constant number (denoted before in the text as $\delta_{_{NLO}}$).
This leads to the VFF high-energy constraints
\begin{eqnarray}
\Frac{F_V  G_V }{F^2}+\hat{\delta}_0 &=& 1+\delta_0\, ,
\nn\\
\widetilde{L}_9 +\hat{\delta}_{2} &=&  0\, .
\end{eqnarray}

Hence, the VFF finally takes the well-behaved structure
(\ref{eq.VFFNLO}) employed in the article,
\begin{eqnarray}
\mF(s) \, &=& \,1\,\,\,+\,\,\,\big(1+\delta_0\big) \, \Frac{s}{M_V^2-s}
\,\,\, +\,\,\,  \sum_{m_1,m_2} \mathcal{F}^{1\ell}(s) |_{m_1,m_2}
\nn\\
&&=\,
\Frac{M_V^2}{M_V^2-s}\,+\,  \mathcal{F}_{_{\rm NLO}}(s)\, .
\end{eqnarray}
Notice that no real double pole term $\hat{\delta}_{-2}$
remains in our perturbative NLO expression as
we have chosen the on-shell mass scheme.

\end{document}